\documentclass[conference]{IEEEtran}
\usepackage{pbox}

\usepackage{cite}

\ifCLASSINFOpdf
  \usepackage[pdftex]{graphicx}
\else
\fi
\usepackage{amsmath}
\usepackage{algorithm}
\usepackage{algorithmic}
\makeatletter
\def\BState{\State\hskip-\ALG@thistlm}
\makeatother

\usepackage{array}
\usepackage{url}

\usepackage{xcolor}
\begin{document}
%
\title{A Deployable Online Optimization Framework for EV Smart Charging with Real-World Test Cases}



%
\author{\IEEEauthorblockN{Nathaniel Tucker\IEEEauthorrefmark{2}
and
Mahnoosh Alizadeh\IEEEauthorrefmark{2}}
\IEEEauthorblockA{\IEEEauthorrefmark{2}Department of Electrical and Computer Engineering, 
University of California, Santa Barbara,
California, 93106, USA}
}


\maketitle


\begin{abstract}
We present a customizable online optimization framework for real-time EV smart charging to be readily implemented at real large-scale charging facilities. Notably, due to real-world constraints, we designed our framework around 3 main requirements. First, the smart charging strategy is readily deployable and customizable for a wide-array of facilities, infrastructure, objectives, and constraints. Second, the online optimization framework can be easily modified to operate with or without user input for energy request amounts and/or departure time estimates which allows our framework to be implemented on standard chargers with 1-way communication or newer chargers with 2-way communication. Third, our online optimization framework outperforms other real-time strategies (including first-come-first-serve, least-laxity-first, earliest-deadline-first, etc.) in multiple real-world test cases with various objectives.  We showcase our framework with two real-world test cases with charging session data sourced from SLAC and Google campuses in the Bay Area.
\end{abstract}


%
\IEEEpeerreviewmaketitle
\makeatletter
\def\blfootnote{\xdef\@thefnmark{}\@footnotetext}
\makeatother

\section{Introduction}
\label{section: introduction}

In the U.S., if federal zero-emission vehicle sales targets are met, there could be more than 48 million electric vehicles (EVs) on the road in 2030 \cite{mckinsey}. In order to provide charge to this growing EV population, it is estimated that over 1.2 million public EV chargers need to be installed at on-the-go locations and at destinations where EVs are parked for long periods \cite{mckinsey}. Furthermore, the estimated cost for hardware, planning, and installation of this future public charging infrastructure exceeds \$35 billion (U.S.D.) \cite{mckinsey}. 

Due to the increasing numbers of EVs, large investment cost of EV charging infrastructure, and the need to charge EVs through cheaper and cleaner energy resources, it is evident that smart charging strategies are required to schedule the power delivery to EVs to maximize the benefits of both the EVs and the public charging infrastructure \cite{ferguson2018optimal, turan_ITSC, moradipari_TTE}. In response, there has been much academic work in the area of EV smart charging at public facilities in recent years \cite{tucker2019online, 8917101, 8635950}; however, most of the effective and implementable solutions require user input, are often facility/infrastructure specific, or ignore critical infrastructure details for modeling purposes. Additionally, the fact that some smart charging frameworks require user input, specifically energy request amounts and departure time estimates, severely constrains their deployment potential. First, most charging infrastructure installed today does not have 2-way communication and would require a third-party user interface (e.g., a touchscreen or smartphone application) to enable user input. Second, the accuracy of user input information is often quite poor. The authors of \cite{lee2019acn, lee_mag} showed that user input data in their real-world smart charging system has 18.6\% and 26.9\% mean absolute error percentages for departure time and energy request estimates, respectively.

Furthermore, while the research area of smart charging has been thoroughly explored in theory, there are only a few frameworks that are actually deployable in the real-world. This is often due to simplifications of the charging infrastructure, assumptions on data availability from the users (i.e., assuming energy request and/or departure times are known a priori), or the lack of real-world data to test and validate the performance of new smart charging strategies. As such, detailed and repeatable case studies for effective and deployable smart charging algorithms could be valuable to the community. 

\textit{Specific Challenges:} There are several key challenges to designing a deployable framework to schedule the charging profiles of numerous EVs. First and foremost, the algorithm must run in real-time without knowledge of the future EV arrivals. The algorithm must adapt its planned power schedules as more information is revealed (i.e., as more EVs arrive to the parking lot). Second, contrary to most smart charging algorithms presented in the literature, if the system is to run on the most common standard chargers under the J1772 standard, the algorithm must be able to function with limited information from each EV \cite{7586042}, \cite{6919255}. Specifically, when an EV plugs in, the algorithm does not get access to the EV's State of Charge (SoC) nor does it know the EV's future departure time (most level 2 chargers do not sense EV SoC nor do they request user input for future departure times). As such, our smart charging algorithm must predict how much energy an EV may consume as well as the EV's future departure time. Such challenges have been acknowledged in past papers including \cite{8585856, 8586075, 8585744, 8586132}. Third, all of the EV charging schedules within a parking lot are coupled due to the local transformer capacity constraint \cite{POWELL2020115352}. As such, the algorithm cannot over-allocate power at any given time; therefore, the algorithm should make use of a model of the future EV arrivals to avoid over allocating power due to unexpected arrivals.

\textit{Contribution:} In this paper, we present a deployable and customizable framework for real-time smart charging to be implemented at real large-scale charging facilities. Three of the main contributions of our framework are as follows:
\begin{enumerate}
    \item  The smart charging strategy is readily deployable and customizable for a wide-array of facilities, infrastructure, objectives, and constraints. 
    \item The online optimization framework can be easily modified to operate with or without user input for energy request amounts and/or departure time estimates which allows our framework to be implemented on standard chargers with 1-way communication or new chargers with 2-way communication.
    \item The online optimization framework  outperforms other real-time strategies (including first-come-first-serve, least-laxity-first, earliest-deadline-first, etc.) in multiple real-world test cases. We showcase our framework with charging session data obtained from collaboration with SLAC and Google campuses \cite{tucker2022real, 9281897} in the Bay Area even with poor accuracy on users' departure time predictions.
\end{enumerate}


\section{Problem Description}
\label{section: problem description}
In the following, we consider an EV charging facility running a real-time smart charging strategy. Specifically, we are examining a time horizon $t=1,\dots,T$ where the facility manager wants to schedule the energy delivery to the EVs that utilize the chargers within the facility. We denote the variable $e_i(t)$ as a $T$x1 vector for the amount of energy (in kWh) that is delivered at each time $t$ to EV $i$. It is through this variable that the smart-charging algorithm can determine when to deliver energy and how much to deliver. Furthermore, throughout the time horizon, we index the EV arrivals as $i=1,\dots,I$ with each EV having an arrival time $t_i^a$, departure time $t_i^d$, and energy demand/request amount $d_i$. We note that the departure time and energy request for each arrival $i$ is unknown unless the facility has communication infrastructure in place to solicit this information from the users. Additionally, we note that any real-time smart charging algorithm must be able to function without knowledge of the arrival sequence a priori. That is, a real-time smart charging algorithm must operate as information is revealed through arrivals and departures within the system. The online optimization framework we present is able to function in real-time with or without the user information for departure times and energy requests, an aspect that is critical for deploying a smart-charging framework to existing charging infrastructure.


\subsection{Objectives}
Due to the fact that we are designing a framework that functions with or without user input (i.e., a framework agnostic to users' energy requests and departure times), we must design an appropriate objective function that is fair to all users, does well in controlling the costs incurred by the facility, and outputs desirable load profiles for the facility.

Accordingly, in this section, we discuss several objectives that the manager of a workplace charging facility might want to optimize. These include maximizing the utility of the EV owners, minimizing Time-Of-Use (TOU) electricity cost, minimizing demand charges, promoting load flattening and equal energy sharing,  or utilizing behind-the-meter renewable generation. Likely, a facility manager would want to consider multiple objectives simultaneously with varying importance for each objective. As such, the main objective function of our optimization framework is a summation of various utility functions multiplied by weights to signify their importance:
\begin{align}
\max_{\substack{e}} U(e) = \max_{\substack{e}} \sum_{f=1}^{F} w_f u_f(e).
\end{align}
Note that $f=1,\dots,F$ are the various utility functions included in the global objective. Additionally, $w_1,\dots,w_F$ are the weights used to tune the global objective and determine the relative importance of each utility function. Now we will discuss several utility functions that could be included by a charging facility operator (some of the following utility functions were adopted from \cite{lee_tsg1, lee_tsg2} which present a similar real-time smart charging algorithm and case studies; however, they make the assumption that user input for energy request amounts and departure times is always available).

\textit{EV Owner Utility:} The main purpose of many workplace charging facilities is to provide utility to the users of the system (e.g., employees or visitors that drive EVs). As such, one objective that is common to many smart charging systems is to maximize user utility based on how much energy their EV receives during their stay (this objective can be used if energy requests are unknown):
\begin{align}
    \label{eq:OU}
    u_{OU}(e) = \sum_{i}\log (\sum_{t}e_i(t)+1).
\end{align}
In \eqref{eq:OU}, we represent the utility of user $i$ as a logarithmic function dependant on how much energy is delivered to user $i$'s EV. The logarithmic utility term was chosen to model the diminishing returns effect in user utility for EVs receiving excessive amounts of energy (e.g., the first 20kWh charged is more valuable to the EV owner than the second 20kWh). We note that this objective does not depend on when the energy is delivered; therefore, if the facility operator wants to emphasize quickly charging the EVs, the following objective can be used.

\textit{Quick Charging:} Here, the facility manager's goal is to deliver as much energy as possible to the EVs (and does not require departure time or energy request information):
\begin{align}
    \label{eq:qc}
    u_{QC}(e) = \sum_{t}\frac{T-t+1}{T}\sum_{i}e_i(t).
\end{align}
In \eqref{eq:qc}, the utility for energy delivered at time $t$ is scaled by a term that decreases as time progresses. That is, energy delivered to EVs in the near future is more valuable than energy delivered later on, thus prioritizing quick charging.

\textit{Maximizing Profit (or Minimizing Cost):} In addition to maximizing user utility, many charging facilities aim to maximize profit or minimize their operational costs. In the following, let $q(t)$ be the price that a user has to pay the facility for 1 kWh of energy at time $t$ and $p(t)$ be the price of electricity from the time-of-use (TOU) rates from the local utility. Furthermore, let us denote $z(t)$ as the energy used by other loads that are not the EVs from behind the same meter (e.g., other buildings, lights, etc). As such, we can write the profit maximization function as follows:
\begin{align}
    u_{PM}(e) = \sum_{t}q(t)\sum_{i} e_i(t) - \sum_{t}p(t)\Big(\sum_{i}e_i(t) + z(t)\Big). 
\end{align}
If $z(t)$ is unknown or unmeasured, it can be set to zero and only the revenue and cost of the EVs will be considered. Furthermore, setting $q(t)=0,\forall t$ allows for the manager to only consider cost minimization.

\textit{Minimizing Demand Charges:} Alongside TOU energy costs, the facility manager needs to be conscious of the maximum demand each month. We denote the monthly demand charge as $\hat{p}$ \$/KW that gets charged based on the peak load each month:
\begin{align}
    u_{DC}(e) = -\hat{p} \cdot \max_{\substack{t}}\big(\sum_{i}e_i(t)+z(t)\big).
\end{align}
This can be a difficult objective to minimize in real-time; therefore, we use an estimate of the peak demand from the previous month, denoted as $\hat{e}_{old}$ and charge the facility operator for any increase $\hat{e}_{inc}$ to the peak load estimate. As such, we adopt the following strategy to include demand charges in our real-time optimization formulation:
\begin{align}
    &u_{DC}(e) = -\hat{p} \cdot \hat{e}_{inc}\\
    &\nonumber \text{where }\\
    &\hat{e}_{inc} = \max_{\substack{t}}\bigg\{\sum_{i}\big(e_i(t)+z(t)\big)-\hat{e}_{old},0\bigg\}.
\end{align}

\textit{Load Flattening:} Another desirable outcome for a facility manager implementing a smart charging algorithm is load flattening. This minimizes the variations in the load from the facility which can help the local utility and reduce the need for extra generation to account for sudden changes:
\begin{align}
    u_{LF}(e) = -\sum_{t} \big( \sum_{i} e_i(t) + z(t) \big) ^2.
\end{align}

\textit{Equal Sharing:} Another useful objective is to promote equal sharing of the facility's resources among the plugged-in EVs. We write this utility function as follows:
\begin{align}
    u_{ES}(e) = -\sum_{t, i} e_i(t)^2.
\end{align}
Additionally, the inclusion of this term in the global objective function ensures a unique optimal solution; specifically, if there are multiple energy delivery schedules that yield the same total utility, then the addition of the equal sharing objective will force the optimal solution to be the one that most evenly distributes energy among the plugged-in EVs.

\textit{Energy Demand: } Lastly, in the offline optimal and online solutions, if users' energy demands are known either from historical data or user input, then the facility tries to deliver a certain amount of energy to each EV denoted as the energy demand $d_i$ of EV $i$. In the offline case, the energy demand can be ensured with a constraint; however, in the online case, to ensure feasibility, sometimes it can be beneficial to include a penalty function for not fulfilling the energy demand. We can denote such a penalty term as follows:
\begin{align}
    \label{eq:ED}
    u_{ED}(e) = -\sum_{i}\bigg(|\sum_{t}e_i(t)-d_i|\bigg).
\end{align}
When the demand for all the EVs is met, then $\sum_{t}e_i(t)=d_i,\forall i$ and the penalty is 0.
\subsection{Constraints}

There are numerous constraints that limit how the smart charging algorithm can distribute energy to the users' EVs:
\begin{subequations}
    \begin{align}
        & \label{c: energy min max}0 \leq e_i(t) \leq e_{max}, \quad&&\forall t,i\\
        & \label{c: energy outside}e_i(t) = 0, \quad&&\forall t\notin [t_i^a,t_i^d]\\
        & \label{c: energy demand}\sum_{t} e_i(t) \leq d_i, \quad&&\forall i\\
        & \label{c: energy trans}\sum_{i} e_i(t) \leq e_{trans}, \quad&&\forall t\\
        &\label{c: demand increase}\hat{e}_{inc}\geq \sum_{i} e_i(t) - \hat{e}_{old}, \hspace{4pt}&&\forall t\\
        &\label{c: demand increase 0}\hat{e}_{inc}\geq0.
    \end{align}
\end{subequations}
The constraints are described as follows: \eqref{c: energy min max} constrains the minimum and maximum amount of energy that a charger can deliver to an EV in one time slot. Constraint \eqref{c: energy outside} ensures that an EV has to be plugged-in for it to receive energy. Constraint \eqref{c: energy demand} ensures that the optimization does not exceed the energy demand of each EV if the energy request amounts for the EVs are known. Constraint \eqref{c: energy trans} is a coupling constraint across all EVs that limits the total energy delivered at any time slot due to the infrastructure limits or the local transformer. Constraints \eqref{c: demand increase} and \eqref{c: demand increase 0} describe the calculation of the incremental increase to the maximum energy demand for demand charge calculation.

\subsection{Offline Optimization}
If the entire sequence of EV arrivals was known for a given time span (i.e., their arrival times, departure times, and energy demands), then one can formulate an offline optimization using the utility functions in \eqref{eq:OU}-\eqref{eq:ED} and constraints \eqref{c: energy min max}-\eqref{c: demand increase 0} to solve for the optimal smart charging strategy for a given time period:
\begin{subequations}
    \begin{align}
        &\nonumber\max_{\substack{e}} U(e)\\
        &\nonumber\text{subject to:}\\
        &\nonumber\eqref{c: energy min max}-\eqref{c: demand increase 0}
    \end{align}
\end{subequations}
However, this is not the case in real-world charging facilities. Instead, information is revealed to the facility manager in an online fashion, and charging decisions must be made without perfect knowledge of the future. As such, in the following section we discuss an online solution akin to model-predictive-control (MPC) that solves the smart charging scheduling problem in real-time. Additionally, it readily handles challenges that come from inaccurate departure time information.

\subsection{Real-Time Smart Charging Algorithm (RTSC-A)}
In this section, we discuss the online optimization framework to solve the smart charging problem in real-time without knowledge of the future arrivals. The approach is akin to model predictive control (MPC) and solves a convex optimization problem at each time step. Specifically, for each time step $t=1,\dots,T$ the optimization solves for the energy output of each EVSE for the next $T-t$ periods. Then, the facility enacts the control strategy for the next time slot, and recomputes the next set of actions at $t=t+1$ (i.e., rolling horizon).

\textit{Departure time scenario generation:} As discussed in the introduction, users' departure time estimates tend to be quite inaccurate or they are unavailable. As such, we generate $n$ potential departure times that the user might leave at (either from historical data or user input, if available), thus allowing the Real-Time Smart Charging Algorithm (RTSC-A) to plan for multiple scenarios. Using these $N$ candidate departure times, we create a scenario in our optimization problem for each potential departure time and solve the optimization across all scenarios. As time progresses, if a potential departure time is no longer feasible (i.e., the potential departure time is the current time step and the EV has not yet departed), then that scenario is removed from the optimization via dynamic scenario weights (let $C_n$ be a weight coefficient for each scenario that is set to 0 if the scenario is no longer feasible). Furthermore let $x_{i,n}$ be the Tx1 binary vector that indicates when EV $i$ is available to charge in scenario $n$.  

\textit{Certainty equivalent control for future EV arrivals:} We make use of our dataset to generate a model for an average day that consists of estimated arrival times, departure times, and energy requests for each day of the week. We then use these daily models in the real-time optimization to account for the unknown future EV arrivals. Specifically, at time $t$, let us assume that there are $J$ EVs in the certainty equivalent daily model that are expected to arrive in the future. Let $x_j$ be the Tx1 binary vector indicating when EV $j$ is available to charge. The decision variables that determine how much energy is delivered at a given time $t$ are Tx1 vectors labeled as $e_1,\dots,e_I$ for the actual EVs plugged in and $e_{I+1},\dots,e_{I+J}$ for the future EV arrivals from the model:

\begin{small}
\begin{subequations}
\begin{align}
    \label{eqn:obj mpc}&\max_{\substack{e}}
    \sum_{i=1}^I \hspace{-1pt}\sum_{n=1}^N \frac{1}{C_n} \Big[U(e_i,x_{i,n})\Big] 
    +\sum_{j=I+1}^{I+J} \Big[U(e_j,x_{j})\Big] 
    \\
    \nonumber&\textrm{subject to:}
    \\
    &0 \leq e_k \leq e_{max}, \hspace{40.5pt} \forall k=1,\dots,I+J
    \label{eqn:0 mpc}
    \\
    &e_i^{\;T} x_{i,n} \leq d_i,
        \label{eqn:deliveredmin mpc actual}
        \;\;\hspace{46pt}\forall i=1,\dots,I,
    \\
        &\nonumber\hspace{23.8ex}\forall n=1,\dots,N,
    \\
        &e_j^{\;T} x_{j} \geq d_j^{min},
        \label{eqn:deliveredmin mpc model}
        \hspace{46pt}\forall j=I+1,\dots,I+J,
    \\
    &\sum_{k=1}^{I+J} e_k(t) \leq e_{trans}, \;\;\hspace{20pt} \forall t=1,\dots,T,
    \label{eqn:transformer}
    \\
    &\hat{e}_{inc}\geq \sum_{k=1}^{I+J} e_k(t) - \hat{e}_{old}, \hspace{7pt}\forall t=1,\dots,T.
    \label{eqn: peak 1 }
\end{align}
\end{subequations}
\end{small}

The first term of the objective function \eqref{eqn:obj mpc} accounts for all $I$ EVs currently plugged in and their $N$ potential departure times each while the second term of the objective function accounts for all $J$ EVs in the future model. Constraint \eqref{eqn:0 mpc} ensures that the energy delivered is nonnegative and less than  the EVSE max $e_{max}$. Constraint \eqref{eqn:deliveredmin mpc actual} ensures that the energy demand of EV $i$ is not exceeded (if energy demand is known), constraint \eqref{eqn:deliveredmin mpc model} ensures a minimum amount of energy is delivered to each EV in the future model, and constraint \eqref{eqn:transformer} ensures that the sum of all energy delivered by the EVSEs at each time $t$ does not exceed the transformer constraint $e_{trans}$. Constraint \eqref{eqn: peak 1 } keeps track of any increase to the current month's peak load for the demand charge. We note that $\hat{e}_{old}$ corresponds to the previous peak energy demand that has been observed during the month. The pseudocode for the daily algorithm can be viewed in Algorithm \ref{algorithm}.

\begin{algorithm}[]
\begin{small}
    \caption{\textsc{Real-Time Smart Charging}}
    \label{algorithm}
    \begin{algorithmic}[1]
    \FOR{ each day}
            \STATE Update current parking lot state 
        \FOR{ each time interval $t$}
            \IF{new departure from parking lot}
                \STATE Update parking lot state
            \ENDIF
            \IF{new arrival to parking lot}
                \STATE Generate $N$ potential departure times for new arrival
                \STATE Update Parking lot state
            \ENDIF
            \STATE \textbf{Formulate optimization for time $t$:}
            \FOR{each EV $i$ plugged in at time $t$}
                \STATE Add EV $i$ to total objective function 
                \eqref{eqn:obj mpc}
                \STATE Add EV $i$ to active constraints \eqref{eqn:0 mpc}-\eqref{eqn: peak 1 }
            \ENDFOR
            \FOR{each future EV $j$ in daily model $t_{model}>t$}
                \STATE Add EV $j$ to total objective function \eqref{eqn:obj mpc}
                \STATE Add EV $j$ to active constraints \eqref{eqn:0 mpc}-\eqref{eqn: peak 1 }
            \ENDFOR
            \STATE \textbf{Solve optimization \eqref{eqn:obj mpc}-\eqref{eqn: peak 1 } for time $t$}
            \STATE Store planned energy schedule for each EV $i$
            \STATE Set each EVSE's output power for the current time interval
            \STATE Update peak load $\hat{e}_{old}$ for demand charge calculation (if a new peak load is observed)
        \ENDFOR
    \ENDFOR
    \end{algorithmic}
\end{small}
\end{algorithm}

\section{Test Cases}
In the following section we compare the performance of our Real-Time Smart Charging Algorithm with various other scheduling strategies. These include the offline optimal (i.e., the optimal schedule solved offline with perfect knowledge of arrival times, departure times, and energy requests), uncontrolled (e.g., First-Come-First-Serve), Least-Laxity-First, and Earliest-Deadline-First. Furthermore, we vary the accuracy of the departure time information that is input to our Real-Time Smart Charging algorithm to showcase its ability to handle inaccurate departure time estimates and still yield good performance. Additionally, we vary the coupling constraint (infrastructure/transformer size) to showcase the performance of our algorithm in a constrained setting. The first test case considers a facility manager that wants to maximize user utility and only slightly cares about TOU electricity costs (i.e., this would be the case of a large company campus who wants to provide free and effective charging for employees). The second test case considers a facility manager that wants to maximize profit while delivering adequate energy to each customer (i.e., this would be the case of a for-profit third-party parking structure equipped with chargers and wants to minimize TOU electricity costs and demand charges). 
\subsection{Test Case Specifics}


\label{section: case study}

We examine a two week period from June 17 - June 29 in 2019 at a Bay Area workplace from our Google EV dataset. The location has 57 level 2 EVSEs  \textcolor{black}{with 50-100 EVs arriving each weekday} and is under PG\&E's E-19 rate structure. 


First, the EV charging session data was filtered by weekday and then filtered again by arrival time. Namely, each charging session was put into one of 12 possible groups corresponding to 2-hour windows for the arrival times (e.g., an EV charging session that started at 9:48am would be stored in the 8:00am-10:00am group). Once this was done, daily arrival time histograms were generated and the average stay duration and average energy consumption were calculated for each of the 12 groups. The average arrivals per weekday, the average arrivals per 2 hour window, the groups' average stay durations, and the groups' average energy consumption were then used to create the algorithm's future model each day and to generate potential departure times for each EV arrival. 
Last, we note that all of these simulations were done in Python with CVX and Mosek on a Laptop with an i7 processor and 16gb of RAM. Moreover, the optimization problem’s complexity is not affected by the number of arriving EVs each day; rather, the problem size grows only as the number of chargers increases. Additionally, for implementation, (13a)-(13f) has to be solved every 15 minutes and for the 57 chargers in our case study, (13a)-(13f) was solved in less than a second. Thus, the algorithm is scalable and there is significant extra time for computation for a larger dataset (i.e., more chargers at the parking lot).



\subsection{Test Case 1: User Utility Maximization with TOU Rates}

\begin{figure*}[]
    \centering
    \includegraphics[width=0.8\textwidth]{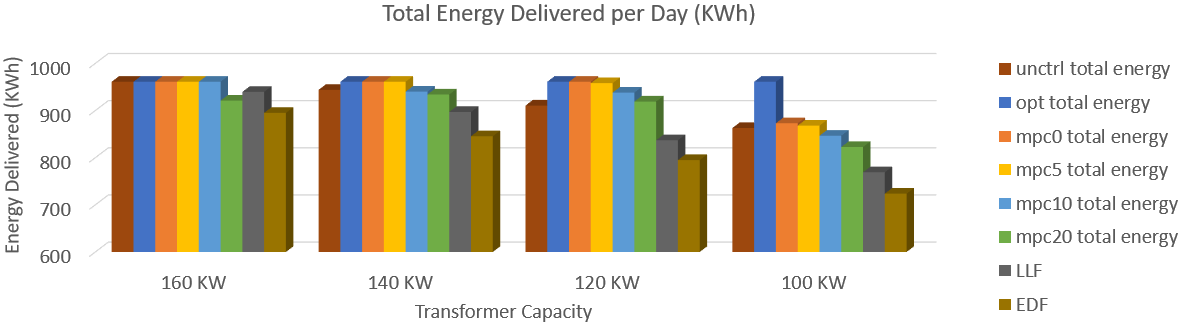}
    \caption{Total energy delivered for the various cases including Least-Laxity-First and Earliest-Deadline-First (both with perfect departure time knowledge) with varying transformer capacities.}
    \label{fig: EM_bars1}
\end{figure*}
\begin{figure}[]
    \centering
    \includegraphics[width=0.9\columnwidth]{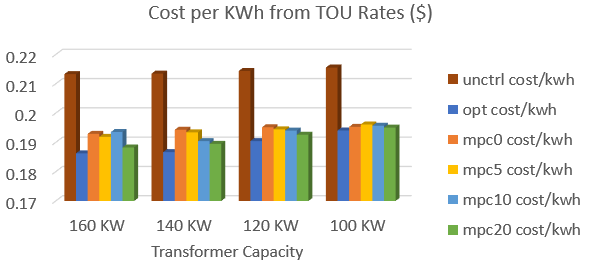}
    \caption{Cost per KWh from TOU rates for the uncontrolled, offline optimal, and 4 MPC test cases}
    \label{fig: EM_bars2}
\end{figure}

In this section, we consider the case of a facility manager that wants to maximize user utility and only slightly cares about TOU electricity costs (i.e., this would be the case of a large company campus who wants to provide free and effective charging for employees). The following objective function was used in the Real-Time Smart Charging Algorithm and the offline optimization:
\begin{align}
    U_{1}(e) =
     15 u_{OU}(e) + u_{PM}(e) + 10^{-9}\Big( u_{LF}(e) + u_{ES}(e)\Big)
\end{align}

In this objective function, the smart charging strategy heavily favors the EV owner's utility with a small emphasis on TOU electricity cost. Load flattening and energy sharing are also included to promote a desirable energy distribution amongst EVs and a flat demand for the utility.

In this test case, we vary the coupling constraint from the local transformer capacity from 160KW to 100KW. We note that in the uncontrolled dataset, the peak load during the two week period was 160.5KW. 

Additionally, we vary the accuracy of the departure time estimates that are input to our RTSC-A. This is due to the fact that users often cannot predict their departure times with perfect accuracy (if the system solicits user information) or due to the fact that historical data does not accurately forecast departure times. As such, for our RTSC-A, for each EV charging session, we consider 10 scenarios. Each of the 10 scenarios makes use of a departure time sampled from a normal distribution centered on the real departure time. We vary the standard deviation of the normal distribution to model EV owners' inaccurate departure time estimates. In the following plots, we denote MPC0 as our RTSC-A with perfect departure time estimates (standard deviation equal to 0) and we denote MPC5 as our RTSC-A with departure time estimates sampled from a normal distribution with standard deviation equal to 5 time steps (1hr 15min). MPC0, MPC5, MPC10, MPC20 make use of normal distributions with standard deviation equal to 0, 5, 10, and 20 time steps, respectively. Furthermore, we note that in the case of MPC20, the standard deviation in the departure time estimate is 20x15min = 5 hours, which is a very inaccurate estimate.

Figure \ref{fig: EM_bars1} presents the total energy delivered to EVs for the various cases including Least-Laxity-First and Earliest-Deadline-First (both with perfect departure time knowledge). Figure \ref{fig: EM_bars2} presents the cost per KWh from TOU rates for the uncontrolled, offline optimal, and 4 MPC test cases.\\
\noindent\textit{Test Case 1 Key Results:}
\begin{enumerate}
    \item As shown in Figure \ref{fig: EM_bars1}, all of the charging strategies were able to deliver ~900KWh of energy per day with a 160KW coupling constraint. However, as the coupling constraint becomes more restrictive (i.e., 140KW, 120KW, or 100KW), many of the charging strategies begin delivering less energy than the offline optimal.
    \item Our RTSC Algorithm with inaccurate departure times consistently beats out EDF and LLF (which are given the exact departure times). This is likely due to the fact that the RTSC algorithm makes use of a future model for arrivals that have not yet shown up, while EDF and LLF are myopic strategies.
    \item As shown in Figure \ref{fig: EM_bars2}, the RTSC algorithm purchases cheaper energy from TOU rates than the uncontrolled case. The uncontrolled case consistently purchases energy at over \$0.21 per KWh for all transformer capacities while the RTSC algorithm remains closer to the offline optimal cost per KWh, below \$0.195 per KWh.
\end{enumerate}

\subsection{Test Case 2: Profit Maximization with TOU Rates and Demand Charges}

\begin{figure}[]
    \centering
    \includegraphics[width=0.95\columnwidth]{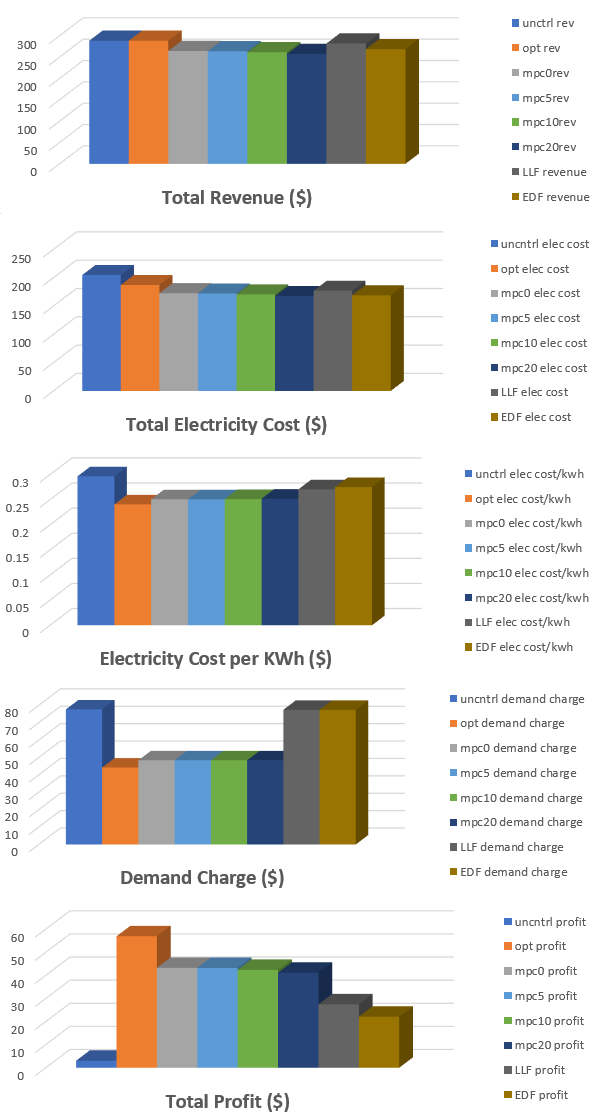}
    \caption{Total revenue, total electricity cost (from both TOU rates and demand charges), electricity cost per KWh, demand charges, and total profit for the uncontrolled, offline optimal, RTSCA with varying departure time accuracies, least-laxity-first, and earliest-deadline-first strategies.}
    \label{fig: PM_bars}
\end{figure}
\begin{figure}[]
    \centering
    \includegraphics[width=0.8\columnwidth]{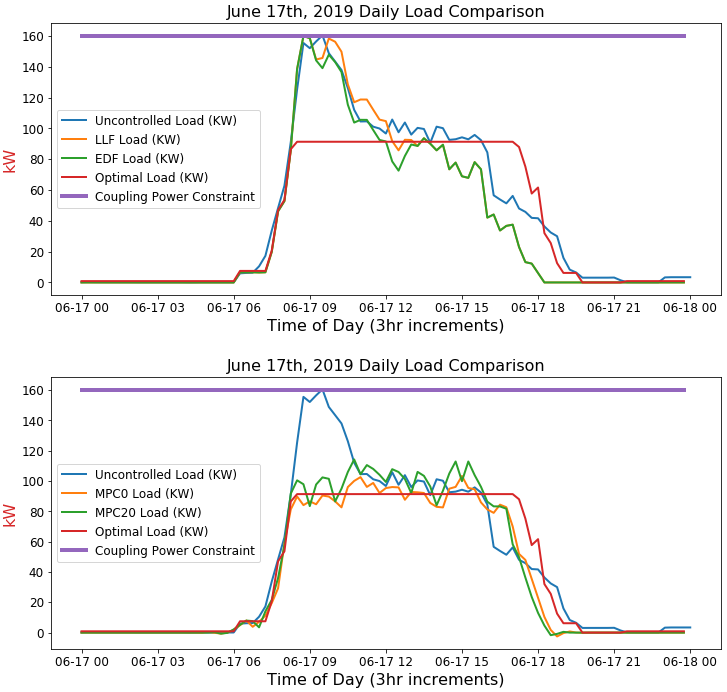}
    \caption{Daily loads of the charging facility for various charging strategies.}
    \label{fig: PM_loads}
\end{figure}

In this section, we consider a facility manager that wants to maximize profit while delivering adequate energy to each customer (i.e., this would be the case of a for-profit third-party parking structure equipped with chargers and wants to minimize TOU electricity costs and demand charges).  The following objective function was used in the Real-Time Smart Charging Algorithm:
\begin{align}
    &U_{2}(e) = \\
    &\nonumber 10 \Big(u_{PM}(e) + u_{DC}(e)\Big)+ u_{OU}(e)  + 10^{-9}\Big( u_{LF}(e) + u_{ES}(e)\Big)
\end{align}

In this objective function, the smart charging strategy prioritizes profit maximization and cost minimization from both TOU rates and demand charges. Load flattening and energy sharing are also included to promote a desirable energy distribution amongst EVs and a flat demand for the utility.

As in the previous test case, we vary the accuracy of the departure time estimates that are input to our RTSC-A. We vary the standard deviation of the normal distribution to model EV owners' inaccurate departure time estimates. Furthermore, in this test case, we do not vary the transformer constraint, it is set to 160KW for all tests. Additionally, we assume energy is sold to the EV users at \$0.30 per kWh.

Figure \ref{fig: PM_bars} presents results for total revenue, total electricity cost (from both TOU rates and demand charges), electricity cost per KWh, demand charges, and total profit for the uncontrolled, offline optimal, RTSC-A with varying departure time accuracies, least-laxity-first, and earliest-deadline-first strategies. Figure \ref{fig: PM_loads} presents the daily load profile of the charging facility for various charging strategies. The top plot includes a comparison of LLF and EDF with the uncontrolled and optimal load. The bottom plot shows the real-time smart charging algorithm's daily load compared to the offline optimal and uncontrolled.

{\textit{Test Case 2 Key Results:}
\begin{enumerate}
    \item As shown in Figure \ref{fig: PM_bars}, total revenue, total electricity cost (from TOU rates and demand charges), and electricity cost per KWh, are similar across all the charging strategies, with the offline optimal performing the best and the RTSC-A with perfect departure time information performing the second best.
    \item However, in plot 4 of Figure \ref{fig: PM_bars}, we see that the demand charges from the uncontrolled case, EDF, and LLF are significantly more than the offline optimal and RTSC-A.
    \item Due to the offline solution's and RTSC-A's ability to optimize for demand charges, the daily net profit (plot 5 in Fig. \ref{fig: PM_bars}) for these strategies significantly outperforms the uncontrolled case as well as EDF and LLF. Even the profit of RTSC-A with bad departure time estimates (RTSC-A with stddev=20) outperforms LLF and EDF with perfect departure time estimates.
    \item RTSC-A's average daily profit is $\sim$\$40, LLF's average daily profit is $\sim$\$23, EDF's average daily profit is $\sim$\$19, the offline optimal average daily profit is $\sim$\$52, and the uncontrolled average daily profit is $\sim$\$3.
    \item As shown in the top plot of Figure \ref{fig: PM_loads}, EDF and LLF load profiles look similar to the uncontrolled profile and reach the coupling constraint of 160KW in the mid-morning. However, in the bottom plot of Figure \ref{fig: PM_loads}, the RTSC algorithm is able to flatten the daily profile to mimic the offline optimal, which is a perfectly flat profile, thus significantly reducing demand charges.
\end{enumerate}}

\section{Conclusion}
In this paper we presented a customizable online optimization framework for real-time EV smart charging to be readily implemented at real large-scale charging facilities. The smart charging strategy is readily deployable and customizable for a wide-array of facilities, infrastructure, objectives, and constraints. The online optimization framework can be easily modified to operate with or without user input for energy requests and/or departure time estimates. Our online optimization framework outperforms other real-time strategies (including first-come-first-serve, least-laxity-first, earliest-deadline-first, etc.) in multiple real-world test cases using real charging session data from SLAC and Google campuses in the Bay Area even with poor accuracy on users' departure time predictions.


\begin{footnotesize}
    \textsc{Acknowledgment:} This work was funded by the California Energy Commission under grant EPC-17-020. SLAC National Accelerator Laboratory is operated for the US Department of Energy by Stanford University  under  Contract  DE-AC02-76SF00515.  Thank you to Rolf Schreiber from Google for providing data.    
\end{footnotesize}



%

\linespread{0.95}
\bibliographystyle{IEEEtran}
\bibliography{references}

\end{document}